\documentclass{elsart5p}
\usepackage{graphics}
\usepackage{graphicx}
\usepackage{amssymb}
\begin{document}

\begin{frontmatter}



\title{On the use of a single site approximation to describe correlation in pure metals}

\author[AA]{C.M. Chaves\corauthref{Name1}},
\ead{cmch@cbpf.br}
\author[AA]{A. A. Gomes},
\author[AA,BB]{A.Troper}

\address[AA]{Centro Brasileiro de Pesquisas F\'\i sicas, Rua Xavier Sigaud 150,Rio de Janeiro, 22290-180, RJ, Brazil}
\address[BB]{Centro Brasileiro de Pesquisas F\'\i sicas, Rua Xavier Sigaud 150,Rio de Janeiro, 22290-180, RJ, Brazil and Universidade do Estado do Rio de Janeiro, Rua S\~ao Francisco Xavier 524, Rio de Janeiro, RJ, Brazil}

\corauth[Name1]{Corresponding author. Tel: (05521) 2141-7285 fax: (05521)
2141-7400}

\begin{abstract}
The magnetic properties of pure transition-like metals are discussed within the single site approximation, to take into account the  electron correlation. The metal is described by two hybridized bands one of which includes the Coulomb correlation. Our results indicate that ferromagnetism follows from adequate values of the correlation and  hybridization. 
\end{abstract}


\end{frontmatter}

\section{Introduction}

The coherent potential approximation (CPA) has been largely used to provide a simple single site description of disordered alloys\cite{general-cpa,shiba}. Another application of CPA procedures is the description of \textit{pure} metals in presence of electron-electron correlations\cite{roth1}. In this formulation, one describes the correlated electron by a spin and energy dependent effective self-energy which incorporates the effects of correlation. This self-energy is self-consistently determined by imposing the vanishing of the scattering $T$ matrix associated to a given site which exhibits the full Coulomb interaction. 

In our model, the metal has two non-degenerate bands, $a$ and $b$. The first is a Hubbard-like narrow band with in-site interaction $U$, hybridized with  the second one, $b$, a broader band. The Hamiltonian we adopt to describe pure metals  is then
\begin{eqnarray}
\label{H}
\mathcal{H}=\sum_{i,\sigma }\epsilon ^{a}a_{i\sigma }^{+}a_{i\sigma
}+\sum_{i,j,\sigma }t_{ij}^{a}a_{i\sigma }^{+}a_{j\sigma
}+\sum_{i,j,\sigma }t_{ij}^{b}b_{i\sigma }^{+}b_{j\sigma
}\\
+\sum_{i}Un_{i\uparrow }^{(a)}n_{i\downarrow }^{(a)}
+\sum_{i,j,\sigma }(V_{ab}b_{i\sigma }^{+}a_{j\sigma }+V_{ba}^{+}a_{i\sigma}^{+}b_{j\sigma })~,\nonumber\
\end{eqnarray}
where  $n_{i\sigma }^{a}=a_{i\sigma }^{+}a_{i\sigma }$ ; $\sigma$ denotes spin. $t_{ij}$ denotes the tunneling amplitudes between neighboring sites $i$ and $j$ , in each band; $\epsilon^{a}$ is the center of the $a$-band.
We follow Roth's approach\cite{roth1} to describe the electron-electron correlation  through an effective Hamiltonian 
\begin{eqnarray}
\label{Heff}
\mathcal{H}_{eff}&=&\sum_{i,\sigma }\epsilon ^{a}a_{i\sigma }^{+}a_{i\sigma
}+\sum_{i,j,\sigma }t_{ij}^{a}a_{i\sigma }^{+}a_{j\sigma
}+\sum_{i,j,\sigma }t_{ij}^{b}b_{i\sigma }^{+}b_{j\sigma
}\\
&+&\sum_{i,\sigma}n_{i\sigma}^{a}\Sigma^{\sigma}+Un^{a}_{0\uparrow }n^{a}_{0\downarrow} 
+\sum_{i,j,\sigma }(V_{ab}b_{i\sigma }^{+}a_{j\sigma }+ \nonumber\
 h.c.)\\ 
&-&\sum_{\sigma}n_{0\sigma}^{a}\Sigma^{\sigma},\nonumber\
\end{eqnarray}
where $\Sigma^{\sigma}$ is the self-energy associated with electron-electron interaction. We recall that the main spirit of the method consists in replacing a tranlationally invariant problem, as defined by (\ref{H}), by the alloy problem, defined by (\ref{Heff}), where only the origin incorporates the Coulomb interaction. The effective Hamiltonian (\ref{Heff}) still includes the difficulty of dealing with the Coulomb intra-atomic term at the origin and we have to resort to some approximation. 

In the Green's function method\cite{zuba,tahir}, the equation of motion of a given Green's function generates higher order  functions and we thus need a decoupling procedure. After some algebra\cite{todos} we end up with the following Green's functions:
\begin{equation}
G^{a}_{kk',\sigma}(w)={\frac{\delta_{kk'}}{w-\tilde{\epsilon}^a_k -\Sigma^\sigma(w)}}, 
\label{gd}
\end{equation}
the Green's function of the bare $b$ band,
\begin{eqnarray}
\label{gsb}
G^{b}_{0,k\sigma}(w)&=&{\frac{1}{w-\epsilon^{b}_k}} ,
\end{eqnarray}
and the Green's function of the renormalized $b$ band
\begin{equation}
\label{gs}
G^{b}_{kk\sigma}(w)=G^{b}_{0,k\sigma}(w)+G^{b}_{0,k\sigma}(w)VG^{a}_{kk,\sigma}(w)VG^{b}_{0,k\sigma}(w). 
\end{equation}
In these equations 
%\begin{equation}
%\epsilon^{a}_{k}={\frac{T_{a}(cos(k_xa)+\cos(k_ya)+\cos(k_za))}{A}}, 
%\label{ed}
%\end{equation}
\begin{equation}
\tilde {\epsilon}^{a}_{k}=\epsilon^{a}_k +{\frac{V_{ab}^2(k)}{w-{\epsilon}^b_k }}, 
\label{etild}
\end{equation}
is the recursion relation of the renormalized  $a$ band; $\epsilon^{a}_k$ and $\epsilon^{b}_k$ denote the bare bands.
The vanishing of the T-matrix gives a self-consistent equation for the self-energy:

\begin{equation}
\label{elivre}
\Sigma^\sigma= U<n^a_{0-\sigma}>+(U-\Sigma^\sigma)F^\sigma(w,\Sigma^\sigma)\Sigma^\sigma,
\end{equation}
with
\begin{equation}
F^\sigma= \sum_k G^{a}_{kk,\sigma}
\label{F}
\end{equation}
This is equivalent to an alloy analogy approximation, where the broadening correction\cite{hubIII} has been neglected.

 Our self-consistent numerical results can describe a transition metal of the iron group (then $a \equiv d$ and $b \equiv s $) or some f electron compounds ($a \equiv f$ and $b \equiv s, p, d $).

\begin{figure}[!ht]
\begin{center}
\includegraphics[angle=0,width=0.45\textwidth]{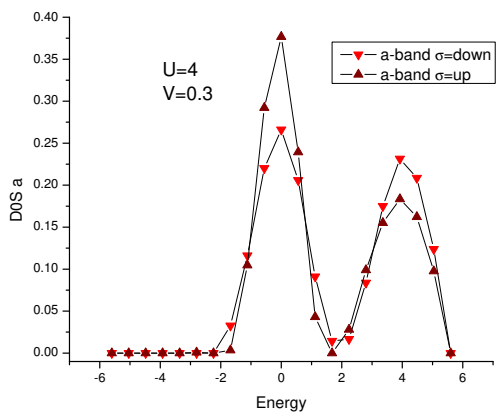}
\end{center}
\caption{Density of states(DOS) of the $a$-band, up and down, for $U=4$ and $V=0.3$. The Fermi level is at $\epsilon_F=0.45$. As a result $n_{\uparrow}^{a}=0.475$ and $n_{\downarrow}^{a}=0.390$, giving a ferromagnetic moment      $m=0.085$. The separation of the centers of the two subbands is $U$, as expected.}
\label{dos}
\end{figure}

 In fig(\ref{dos}) we present the density of states (DOS) of the $a$-band for $U=4$ and $V=0.3$. In same units, the bare $a$ band width is $W=2$. The Fermi level is  at $\epsilon_F=0.45$. This gives $n_{\uparrow}^{a}= 0.475$ and $n_{\downarrow}^{a}=0.390$ and thus a ferromagnetic moment of $m_{a}=0.085$ is developed. The DOS here obtained exhibits a bimodal structure caracterizing a Hubbard strongly correlated regime.

\begin{figure}
\begin{center}
\includegraphics[angle=0,width=0.45\textwidth]{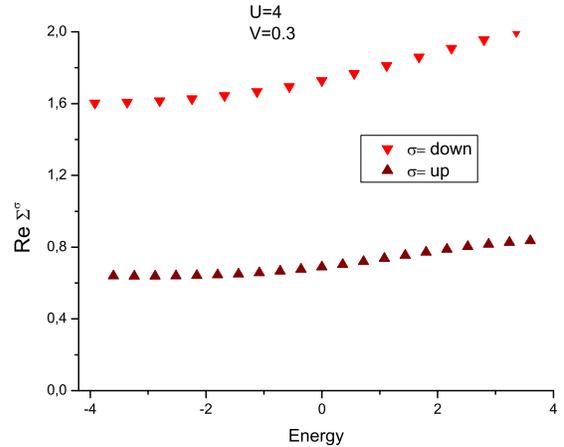}
\end{center}
\caption{Real part of the self-energy for up and down spins for $U=4$ and $V=0.3$.}\label{sigma}
\end{figure}

In fig (\ref{sigma}) the real part of $\Sigma^{\sigma}$ is displayed for $U=4$ and $V=0.3$. $\Sigma^{\sigma}$ is a smooth funcion of energy and for this $U$ differs considerably from the constant Hartree-Fock values $(\Sigma^{\uparrow}=1.56 , \Sigma^{\downarrow}=1.90)$.

\section{Summary}
The traditional view of the origin of ferromagnetism in metals has been under intense scrutiny recently \cite{bat,voll}). Conventional mean-field calculations favor ferromagnetism but corrections tend to reduce the range of validity of that ground state \cite{voll}. In this paper, using the single site approximation, we obtain ferromagnetic solution for a set of parameters (e.g. $U/W=2$ and $V/W=0.15$). A more complete study bringing up the interplay between $U$ and $V$ will be published elsewhere.

\section{Acknowledgement}
CMC and AT aknowledge the support from the brazilian agencies $CNP_q$ and $PCI/MCT$.

\end{document}